\documentstyle[twoside,12pt]{article}

\newtheorem{Sec3}{3.}
\begin{document}
\setcounter{equation}{0}
\setcounter{section}{0}
\title{\bf On the differential invariants of a family of diffusion equations}
\author{M. Torrisi \footnote{Author to whom correspondence should be addressed}, R. Tracin\`a \\
{\small \it Department of Mathematics and Computer Science}\\
{\small \it University of Catania}\\
{\small \it Viale A. Doria, 6,  95125 Catania,  Italy}\\
{\small \it E-mail: torrisi@dmi.unict.it and tracina@dmi.unict.it}
}

\maketitle

\begin{abstract}
The equivalence transformation  algebra $L_{\cal E}$
for the class of equations $u_t -u_{xx}=f(u,\, u_x) $ is obtained.
After getting the differential invariants with respect to $L_{\cal E}$,
some results which allow to linearize a subclass of equations are showed.
Equations like the standard deterministic KPZ equation fall in this subclass.
\end{abstract}

 \textit{PACS:} 02.30.Jr; 02.20.Tw\\

 \textit{Keywords:} Nonlinear diffusion equations, equivalence
 transformation, differential invariants.\\[2ex]

\newpage
\section{Introduction}
\setcounter{equation}{0}
In some previous papers \cite{Trac1}, \cite{Trac2}, \cite{TTV} the differential
invariants for the family of equations $u_{tt}-u_{xx}=f(u,\,u_t,\,u_x)$ have been obtained and
applied in order to characterize some linearizable subclasses of that
equations. 

Here, we consider the following diffusion equations:
\begin{equation}
u_t -u_{xx}=f(u,\, u_x),  \label{eq1}
\end{equation}
which arise in several problems of mathematical physics.

By using the invariance Lie infinitesimal criterion \cite{OV}, we construct the 
algebra $L_{\cal E}$ of the equivalence transformations.
These transformations have the property to change any element of a family
of PDEs to a PDE which belongs to the same family.
An equivalence transformation maps solutions of an equation of
the family to solutions of the transformed equation.

Following the method proposed by N.H. Ibragimov in \cite{NI}, \cite{NI1} and successively
applied in \cite{ITV}, 
we calculate the differential invariants with respect to the
equivalence transformations of the family (\ref{eq1}).

Starting from these results, we characterize a subclass
of equations (\ref{eq1}) which can be linearized through an equivalence transformation.
In this subclass falls the standard deterministic Kardar-Parisi-Zhang (KPZ) equation \cite{KPZ}, \cite{KS}
which models the relaxation of an initially rough surface
to a flat surface.

The outline of the paper is the following. In Section 2,
 we obtain the infinitesimal equivalence generator of equations (\ref{eq1}).
In Section 3, we look for differential invariants
and, by following the infinitesimal method \cite{NI}, \cite{NI1},
we show that the family of equations (\ref{eq1}) does not admit differential
invariants of order zero and one,
while second order differential invariants
are found. 
Finally, in Section 4, these last ones are used in order to 
characterize a subclass of the family (\ref{eq1}) which can be mapped, by
an equivalence transformation, in the Fourier's equation.
The conclusions are given in Section 5.

\section{Algebra $\mathbf{L_{\cal E}}$}
\setcounter{equation}{0}
We recall that a transformation of the type
\begin{eqnarray}
t = t (\tau ,\,\sigma ,\, v),\quad
 x = x (\tau ,\,\sigma ,\, v),\quad
u = u(\tau ,\,\sigma ,\, v),\label{trasformazione3} 
\end{eqnarray}
which is locally a $C^{\infty}$-diffeomorphism and
changes the original equation into a new equation having
the same differential structure but a different form of the function $f$, is
 an equivalence transformation \cite{OV} 
(hereafter ET) for the equations (\ref{eq1}).
An invariance transformation can be regarded as particular
ET such that the transformed function $f$ has the same form.
In the following we consider only continuous groups of equivalence transformations.
\par
The direct search for the equivalence transformations through the finite
form of the transformation is connected with 
considerable computational difficulties. The use of the Lie infinitesimal criterion, 
suggested by Ovsiannikov \cite{OV}, gives an algorithm to find 
the infinitesimal 
generators of the ETs that overcame these problems.

In order to obtain a continuous group of ETs of equations 
(\ref{eq1}),
we consider, by following \cite{OV}, the arbitrary 
function $f$ as a dependent variable and apply the  
Lie  infinitesimal invariance
criterion  to the following system:
\begin{eqnarray}
u_t -u_{xx}=f, \nonumber\\
f_t=f_x=f_{u_t}=0,  \label{condaux}
\end{eqnarray}
where the last three equations of (\ref{condaux})
are usually called {\it auxiliary equations} and
 give the independence of $f$ on $t$, $x$ and $u_t$.

The infinitesimal equivalence generator  $Y$ has  the following form:
\begin{eqnarray}
Y = \xi^1 \frac{\partial}{\partial t} + \xi^2 \frac{\partial}{\partial x}
+ \eta \frac{\partial}{\partial u} + \zeta_1\frac{\partial}{\partial u_t}
+ \zeta_2\frac{\partial}{\partial u_x} + \mu \frac{\partial}{\partial f}, \label{Y}
\end{eqnarray}
where $\xi^1$, $\xi^2$ and $\eta$ are sought depending on $t$, $x$ and
$u$, while $\mu$ depends on  $t$, $x$, $u$, $u_t$,
$u_x$ and $f,$   the components $\zeta_1$ and $\zeta_2$, as known, are given by
\begin{eqnarray}
\zeta_1=D_t(\eta)-u_tD_t(\xi^1)-u_xD_t(\xi^2), \quad
 \zeta_2=D_x(\eta)-u_tD_x(\xi^1)-u_xD_x(\xi^2).
\end{eqnarray}
The operators $D_t$ and $D_x$ denote the total derivatives
with respect to $t$ and $x$:
\begin{eqnarray}
& & D_t = \frac{\partial}{\partial t} + u_t \frac{\partial}{\partial u} + u_{tt} \frac{\partial }{\partial u_t} +
u_{tx} \frac{\partial}{\partial u_x} + ... , \\
& & D_x = \frac{\partial}{\partial x} + u_x \frac{\partial}{\partial u} + u_{tx} \frac{\partial}{\partial u_t} +
u_{xx}\frac{\partial}{\partial u_x} + ... .
\end{eqnarray}
The prolongation of operator (\ref{Y}),  which we need  in order
to require the invariance of  (\ref{condaux}), is
\begin{eqnarray}
\widetilde Y =Y + \zeta_{22} \frac{\partial}{\partial u_{xx}}+\omega _t\frac{\partial}{\partial f_t}+
\omega _x \frac{\partial}{\partial f_x}+\omega _{u_t} \frac{\partial}{\partial f_{u_t}}, \label{Yprol}
\end{eqnarray}
where (see e.g. \cite {ITV}, \cite {ITV1})
\begin{eqnarray}
&&\zeta_{22}=D_x(\zeta_2)-u_{tt}D_x(\xi^1)-u_{xx} D_x(\xi^2),\\
&&\omega _t=\widetilde D_t (\mu)-f_u \widetilde D_t (\eta)-f_{u_x}\widetilde D_t (\zeta_2),\label {omegat}\\
&&\omega _x=\widetilde D_x (\mu)-f_u \widetilde D_x (\eta)-f_{u_x}\widetilde D_x (\zeta_2),\label {omegax}\\
&&\omega _{u_t}=\widetilde D_{u_t} (\mu)-f_u \widetilde D_{u_t} (\eta)-f_{u_x}\widetilde D_{u_t} (\zeta_2),\label {omegaut}
\end{eqnarray}
while $\widetilde D_t$,  $\widetilde D_x$ and $\widetilde D_{u_t}$
are defined by:
\begin{eqnarray}
& & \widetilde D_t = \frac{\partial}{\partial t} + f_t
\frac{\partial }{\partial f} + f_{tt}
\frac{\partial }{\partial f_t} + f_{tx} \frac{\partial }{\partial f_x} + ... , \label {Dtildet}\\
& & \widetilde D_x = \frac{\partial}{\partial x} + f_x
\frac{\partial }{\partial f} + f_{tx} \frac{\partial
}{\partial f_t} + f_{xx} \frac{\partial }{\partial f_x} +
... .,\label {Dtildex}\\
& & \widetilde D_{u_t} = \frac{\partial}{\partial {u_t}} + f_{u_t}
\frac{\partial }{\partial f} + f_{t u_t} \frac{\partial
}{\partial f_t} + f_{xu_t} \frac{\partial }{\partial f_x} +
... .\label {Dtildeut}
\end{eqnarray}
Applying the operator (\ref{Yprol}) to the system (\ref{condaux}) and
following the well known algorithm (see e. g. \cite{ITV1}, \cite{TTV1}) we obtain
\begin{eqnarray}
Y&=&\left(c_0+c_1 t\right)\frac{\partial}{\partial t}+\left(\frac{1}{2}c_1 x+c_2 t+ c_3\right)\frac{\partial}{\partial x}+
\varphi (u)\frac{\partial}{\partial u}+ \nonumber \\
&&+\left(-c_1 u_t-c_2 u_x+\varphi ' u_t\right)\frac{\partial}{\partial {u_t}}+
\left(-\frac{1}{2}c_1 u_x+\varphi 'u_x\right)\frac{\partial}{\partial {u_x}}+ \nonumber \\
&&+\left(-c_1 f-c_2 u_x+\varphi ' f-\varphi '' u_x^2\right)\frac{\partial}{\partial f},\label{gen}
\end{eqnarray}
where $c_0,\,c_1,\,c_2$ and $c_3$ are arbitrary constants,  $\varphi $ is an arbitrary function of $u$
and  the prime denotes the differentiation with
respect to $u$.
So, we have found that the Lie algebra $L_{\cal E}$ for
 the class of equations (\ref{eq1}) is infinite-dimensional and generates
 an infinite continuous group $G_{\cal E}$ of equivalence
transformations spanned
by the following operators:

\begin{eqnarray}
&& {Y}_0=\frac{\partial}{\partial t},\quad\quad
 {Y}_1=t \frac{\partial}{\partial t}+\frac{1}{2}\,x\frac{\partial}{\partial x}-f\frac{\partial}{\partial f}-
u_t\frac{\partial}{\partial {u_t}}-\frac{1}{2}\,u_x \frac{\partial}{\partial {u_x}},\nonumber\\
&& {Y}_2=t \frac{\partial}{\partial x}-u_x \frac{\partial}{\partial f}-u_x \frac{\partial}{\partial {u_t}},\quad\quad
 {Y}_3 = \frac{\partial}{\partial x},\nonumber\\
&&{Y}_\varphi = \varphi  \frac{\partial}{\partial u}+\left(\varphi ' f-\varphi '' u_x^2\right)  \frac{\partial}{\partial f}+
\varphi ' u_t  \frac{\partial}{\partial u_t}+\varphi ' u_x  \frac{\partial}{\partial u_x}.\nonumber
\end{eqnarray}

\section {Search for differential invariants }
\setcounter{equation}{0}
Following \cite {NI}-\cite {ITV}, we recall that, for the family of equations (\ref{eq1}), a {\it differential invariant of order s} is
a function $J$,  of
the independent variables $t$, $x$, the dependent variable $u$ and its derivatives
$u_t$, $u_x$, as well as of the function $f$ an its derivatives of maximal order {\it s}, 
invariant with respect to the equivalence group $G_{\cal E}$ . 
\vskip 5mm
\begin{Sec3}
{\bf Differential invariants of order zero.}
\end {Sec3}
Here we search for functions
\begin{equation}
J = J(t, x, u, u_t, u_x, f) \label{INV0}
\end{equation}
satisfying the invariant condition $Y(J)=0$.
\par
>From the invariant tests $Y_0(J) = 0$ and $Y_3(J) = 0$,
easily follows  that $J$ must depend only on $u$, $u_t$, $u_x$ and $f$.
\par
>From invariance test $ Y_{\varphi} (J)=0$,
after  observing that, being $\varphi$ an arbitrary function,  $Y_\varphi $ can be splitted  in the following
three operators
$$ {\hat Y}_\varphi=\frac{\partial}{\partial u},\quad
 {\hat Y}_{\varphi'}=f \frac{\partial }{\partial f}+u_t  \frac{\partial }{\partial u_t}+u_x  \frac{\partial }{\partial u_x},\quad
 {\hat Y}_{\varphi''}=-u_x^2\frac{\partial}{\partial f},$$
 it is a simple matter to get
\begin{equation}
J=J(q),
\end{equation}
with
\begin{equation}
q=\frac{u_t}{u_x}.
\end{equation}
>From  $Y_2(J) = 0$ we get $J_q=0$, hence   the equations (\ref{eq1}) do not possess differential invariants of zero order.
\vskip 5mm
\begin{Sec3}
{\bf Differential invariants of first order.}
\end {Sec3}
The differential invariants of first order  involve $ f_u $ and $f_{u_x} $ also,
 so we need
the following first prolongation of operator $Y$:
\begin{eqnarray}
Y^{(1)} &=&Y+\omega _u \frac{\partial }{\partial f_u}+ 
\omega _{u_x} \frac{\partial }{\partial f_{u_x}}, \label{prolY}
\end{eqnarray}
which, after computing $\omega _u$ and  $ \omega _{u_x}$  likewise (\ref{omegat})- (\ref{Dtildeut}), can be written as:
\begin{eqnarray}
Y^{(1)} &=&Y+\left( -c_1f_u+\varphi '' f-\varphi '' u_x f_{u_x}-\varphi ''' u_x^2\right) \frac{\partial }{\partial f_u}+ \nonumber \\
&&-\left( \frac{1}{2}\,c_1 f_{u_x}+c_2+2\varphi '' u_x\right) \frac{\partial }{\partial f_{u_x}}. \label{Y1}
\end{eqnarray}
By observing that:
$${Y}^{(1)}_0={Y}_0,\quad{Y}^{(1)}_3={Y}_3\quad {\hat Y}^{(1)}_\varphi= {\hat Y}_\varphi $$
it is a simple matter to ascertain that at this step the search is reduced to look 
for invariant functions of the form
\begin{equation}
J = J(u_t, u_x, f, f_{u},f_{u_x}) 
\end{equation}
with respect to the following operators
\begin{eqnarray}
{Y}^{(1)}_1&=&t \frac{\partial}{\partial t}+\frac{1}{2}x\frac{\partial}{\partial x}-f\frac{\partial}{\partial f}-
u_t\frac{\partial}{\partial {u_t}}-\frac{1}{2}u_x \frac{\partial}{\partial {u_x}}-f_u \frac{\partial}{\partial f_u}+\nonumber \\
&&-\frac{1}{2}f_{u_x}\frac{\partial}{\partial f_{u_x}},\\
{Y}^{(1)}_2&=&t \frac{\partial}{\partial x}-u_x \frac{\partial}{\partial f}-u_x \frac{\partial}{\partial {u_t}}
-\frac{\partial}{\partial f_{u_x}},\\
{\hat Y}^{(1)}_{\varphi '} &=& f \frac{\partial}{\partial f}+u_t  \frac{\partial}{\partial u_t}+ u_x  \frac{\partial}{\partial u_x},\\
{\hat Y}^{(1)}_{\varphi ''}&=&- u_x^2  \frac{\partial}{\partial f}+\left(f-u_x f_{u_x}\right) \frac{\partial}{\partial f_u}
-2u_x \frac{\partial}{\partial f_{u_x}},\\
{\hat Y}^{(1)}_{\varphi '''}&=&- u_x^2  \frac{\partial}{\partial f_u}.
\end{eqnarray}
By requiring the invariance  of  $J$  with respect to the operator  ${\hat Y}^{(1)}_{\varphi '''}$   it follows
\begin{equation}
J = J(u_t, u_x, f, f_{u_x}), \label{INV11}
\end{equation}
while the invariant test, applied to (\ref {INV11})
\begin{equation}
 \hat Y_{\varphi ''}^{(1)}(J) \equiv u_x \,\frac{\partial J}{\partial f} +2 \,\frac{\partial J}{\partial f_{u_x}} = 0,
\end{equation}
yields that 
\begin{equation}
J = J\left(u_t, u_x, p_1\right), \label{INV12}
 \end{equation}
where
\begin{equation}
p_1= \frac{f}{u_x}-\frac{f_{u_x}}{2}.
 \end{equation}
Acting by operator $\hat Y_{\varphi '}^{(1)} $  on the invariant (\ref{INV12}), one
obtains that
\begin{equation}
\hat Y_{\varphi '}^{(1)}(J) \equiv  u_t \,\frac{\partial J}{\partial u_t} +u_x \,\frac{\partial J}{\partial u_x} = 0
\end{equation}
 and hence
\begin{equation}
J = J\left(p_1,\, p_2\right), \label{INV13}
 \end{equation}
with
\begin{equation}
p_2= \frac{u_t}{u_x}.
 \end{equation}
Finally, from the invariant test $Y_2^{(1)}(J)=0$ we get
\begin{equation}
J = J(p), \label{INV14}
 \end{equation}
where
\begin{equation}
p= p_1-2p_2=\frac{f-u_x\,f_{u_x}}{2u_x}-2\frac{u_t}{u_x},
 \end{equation}
and from $Y_1^{(1)}(J)=0$ it follows
\begin{equation}
Y_1^{(1)}(J) \equiv  \left(-\frac{1}{2}\frac{f}{u_x}+ \frac{u_t}{u_x}+\frac{1}{4}f_{u_x}\right)\,\frac{\partial J}{\partial p} = 0.
\end{equation}
Provided that
 $\displaystyle -\frac{1}{2}\frac{f}{u_x}+ \frac{u_t}{u_x}+\frac{1}{4}f_{u_x}\not= 0$,
we get $\frac{\partial J}{\partial p} = 0$. 
Hence the equations  (\ref{eq1}) do not admit differential invariants of first order.
\vskip 5mm
\begin{Sec3}
{\bf Differential invariants of second order.}
\end {Sec3}
In the search for  second order differential invariants, because of
the function $J$ is sought as depending from $f_{uu}$, $f_{uu_x}$ and $f_{u_x u_x}$ also,
we need
the following second prolongation of operator $Y$:
\begin{eqnarray}
Y^{(2)} = Y^{(1)} + \omega _{uu}\frac{\partial }{\partial f_{uu}} +\omega _{uu_x}
\frac{\partial }{\partial f_{uu_x}}+\omega _{u_xu_x}\frac{\partial }{\partial f_{u_xu_x}} \label{2Y}
\end{eqnarray}
where  \cite {ITV}, \cite {ITV1}:
\begin{eqnarray}
&\omega_{uu}&= \widetilde D_u(\omega_u) - f_{uu} \widetilde
D_u(\eta) - f_{uu_x} \widetilde D_u(\zeta_2),\\
&\omega_{uu_x}&= \widetilde D_{u_x}(\omega_u) - f_{uu} \widetilde
D_{u_x}(\eta) - f_{uu_x} \widetilde D_{u_x}(\zeta_2),\\
&\omega_{u_xu_x}&= \widetilde D_{u_x}(\omega_{u_x}) - f_{uu_x} \widetilde
D_{u_x}(\eta) - f_{u_xu_x} \widetilde D_{u_x}(\zeta_2).
\end{eqnarray}
After some calculations we get
\begin{eqnarray}
Y^{(2)} &&=Y^{(1)}-\left(\varphi 'f_{u_xu_x}+2\varphi ''\right) \frac{\partial }{\partial f_{u_xu_x}}+\nonumber\\
&&-\left( \frac{1}{2}\,c_1 f_{uu_x}+\varphi 'f_{uu_x}+\varphi '' u_xf_{u_xu_x}-2\varphi '''u_x\right)
 \frac{\partial }{\partial f_{uu_x}}+\\
&&+\left[ -c_1f_{uu}-\varphi ' f_{uu}+\varphi ''\left( f_u -2 u_x f_{uu_x}\right)
+\varphi ''' (f-u_xf_{u_x})-\varphi ^{\small \mathit IV}u_x^2\right] \frac{\partial }{\partial f_{uu}}. \nonumber\label{Y2}
\end{eqnarray}
By observing that:
$${Y}^{(2)}_0={Y}_0,\quad{Y}^{(2)}_3={Y}_3\quad {\hat Y}^{(2)}_\varphi= {\hat Y}_\varphi $$
we ascertain that we  must look 
for invariant functions of the form
\begin{equation}
J = J(u_t,u_x,f,f_u,f_{u_x},f_{uu},f_{uu_x},f_{u_xu_x}) \label{INV2}
\end{equation}
which are invariant with respect to the following operators:
\begin{eqnarray}
{Y}^{(2)}_1&=&{Y}^{(1)}_1-f_{uu} \frac{\partial}{\partial f_{uu}}-
\frac{1}{2}f_{uu_x} \frac{\partial}{\partial f_{uu_x}},\\
{Y}^{(2)}_2&=&{Y}^{(1)}_2, \\
{\hat Y}^{(2)}_{\varphi '} &=& {Y}^{(1)}_{\varphi '} -f_{uu} \frac{\partial}{\partial f_{uu}}-f_{uu_x} \frac{\partial}{\partial f_{uu_x}}
-f_{u_xu_x} \frac{\partial}{\partial f_{u_xu_x}},\\
{\hat Y}^{(2)}_{\varphi ''}&=&{Y}^{(1)}_{\varphi ''}+\left(f_u-2u_xf_{uu_x}\right) \frac{\partial}{\partial f_{uu}}-
u_x f_{u_xu_x} \frac{\partial}{\partial f_{uu_x}}-2 \frac{\partial}{\partial f_{u_xu_x}},\\
{\hat Y}^{(2)}_{\varphi '''}&=&{Y}^{(1)}_{\varphi '''}+\left(f-u_x f_{u_x}\right)\frac{\partial}{\partial f_{uu}}
-2u_x\frac{\partial}{\partial f_{uu_x}},\\
{\hat Y}^{(2)}_{\varphi ^{IV}}&=&-u_x^2\frac{\partial}{\partial f_{uu}}.
\end{eqnarray}
The invariance condition ${\hat Y}^{(2)}_{\varphi ^{IV}}(J)=0$
implies
\begin{equation}
J = J(u_t, u_x, f, f_u, f_{u_x},  f_{uu_x}, f_{u_xu_x} ) \label{INV21}
\end{equation}
while ${\hat Y}_{\varphi '''}^{(2)} (J) = 0$ yields
\begin{equation}
J = J(u_t, u_x, f,  f_{u_x},  f_{u_xu_x}, p_1), \label{INV22}
\end{equation}
where
\begin{equation}
p_1= \frac{f_u}{u_x}-\frac{f_{uu_x}}{2} .
\end{equation}

Likewise, from $Y_2^{(2)} (J) = 0$ we obtain
\begin{equation}
J = J(u_x,  f_{u_xu_x}, p_1, p_2, p_3 ), \label{INV23}
\end{equation}
where
\begin{equation}
p_2= f-u_t, \quad\quad p_3= \frac{f}{u_x}-f_{u_x},
\end{equation}
and from  $\hat Y_{\varphi '}^{(2)} (J) = 0$ , 
\begin{equation}
J = J(q_1, q_2, q_3, q_4 ), \label{INV24}
\end{equation}
where
\begin{eqnarray}
q_1&=&f_{u_xu_x}(f-u_t),\quad q_2=\left(\frac{f_u}{u_x}-\frac{f_{uu_x}}{2}\right)\left(f-u_t\right),\label{q1}\\
q_3&= &p_3=\frac{f}{u_x}-f_{u_x},\quad  q_4=u_x f_{u_xu_x}\label{q2}.
\end{eqnarray}
By applying the operator $\hat Y_{\varphi ''}^{(2)} $ to the differential invariant given by (\ref{INV24}), taking into account
(\ref{q1}-\ref{q2}), we get  
\begin{equation}
\left(-2\frac{q_1}{q_4}-q_4\right)\frac{\partial J}{\partial q_1}+
\left(\frac{1}{2}q_1-\frac{q_2q_4}{q_1}+\frac{q_1q_3}{q_4}\right)\frac{\partial J}{\partial q_2}
+\frac{\partial J}{\partial q_3}-2\frac{\partial J}{\partial q_4}=0.
\end{equation}
The corresponding characteristic equations give
\begin{equation}
J = J(r_1, r_2, r_3), \label{INV25}
\end{equation}
where
\begin{eqnarray}
r_1&=&\frac{f}{u_x}-f_{u_x}+\frac{1}{2}u_x f_{u_xu_x},\label{r1}\\
r_2&=&f_u-\frac{1}{2}u_x f_{uu_x}+\frac{1}{2}f f_{u_xu_x}-\frac{1}{2}u_x f_{u_x} f_{u_xu_x}+
\frac{1}{4}u_x^2 f^2_{u_xu_x},\label{r2}\\
r_3&=&\frac{f-u_t}{u_x}-\frac{1}{2}u_x f_{u_xu_x}.\label{r3}
\end{eqnarray}
Finally, the invariant test
\begin{equation}
Y_1^{(2)}(J) = 0,
\end{equation}
after some calculations, yields
\begin{equation}
r_1 \frac{\partial J}{\partial r_1}+ 2r_2\frac{\partial J}{\partial r_2}+r_3\frac{\partial J}{\partial r_3}= 0.
\end{equation}
>From the   corresponding characteristic equations, provided that
\begin{equation}
2f-2u_t-u_x^2f_{u_xu_x}\not= 0,
\end{equation}
we get that the general form of second order differential invariants of equation (\ref {eq1}) is
\begin{equation}
J = J(\lambda _1,\lambda _2), \label{INV26}
\end{equation}
with $\lambda _1$ and $\lambda _2$ given by
\begin{eqnarray}
\lambda _1&=&{\displaystyle\frac{2f-2u_xf_{u_x}+u_x^2f_{u_xu_x}}{2f-2u_t-u_x^2f_{u_xu_x}}},\label{lambda1}\\
\lambda _2&=&{\displaystyle\frac{\left(4f_u-2u_xf_{uu_x}+2ff_{u_xu_x}-2u_xf_{u_x}f_{u_xu_x}+
u_x^2f^2_{u_xu_x}\right) u_x^2}{(2f-2u_t-u_x^2f_{u_xu_x})^2}}.\label{lambda2}
\end{eqnarray}

\section{Some Applications}
\setcounter{equation}{0}
Here we wish use  the second order invariants  $\lambda_1$ and  $\lambda_2$
in order to bring  nonlinear equations of the class (\ref {eq1}) in linear form by using the equivalence transformations 
 of the admitted 
group $G_{\cal E}$.
\par
The search for transformations mapping  a non linear differential equation in a linear differential equation
has interested several authors. In particular S. Kumei and G. W. Bluman in their pionering paper   \cite{KB}
gave some necessary and sufficient conditions that, by examining the invariance algebra, allow to affirm whether 
 a nonlinear equation is trasformable in linear form. 
\par
It is worthwhile  noticing that  the Kumei-Bluman algorithm (see also \cite{BK}) constructing the linearizing map, 
based on the existence of 
an admitted infinite parameter Lie group transformations, does not require the knowledge, {\it a priori}, of a specific 
linear target 
equation. The target come out in a natural way during the developments of the algorithm.
Here, instead,  we search the nonlinear equations
of the class (\ref {eq1}) that can be mapped by an equivalence transformation
in a linear equation of the subclass
\begin{eqnarray}
v_\tau  -v_{\sigma \sigma }=k_0 v_\sigma ,  \label{linear}
\end{eqnarray}
with $k_0=const.$
\par
That is, once fixed {\it a priori} the  target  (\ref {linear}) we  characterize the whole set of equations  (\ref {eq1}) which 
can be mapped in (\ref {linear}).
\par
For the subclass  (\ref {linear}) the differential invariants $\lambda _1$ and $\lambda _2$
are zero.
So, taking into account  (\ref {lambda1}- \ref {lambda2}),
we search the functional forms of $f(u,u_x)$ for which 
\begin{eqnarray}
\left \{
\begin{array}{l}
 \lambda _1=0 \\
 \lambda _2=0.
\end{array}
\right.
\end{eqnarray}

Then, solving
\begin{eqnarray}
2f-2u_xf_{u_x}+u_x^2f_{u_xu_x}=0,\label{cond1}
\end{eqnarray}
we get
\begin{eqnarray}
f=u^2_x\,h(u)+h_1(u) u_x  \label{formaf0}
\end{eqnarray}
where $h$ and $h_1$ are arbitrary functions of $u$.

By requiring that
\begin{eqnarray}
\lambda _2\left.\right|_{f=u^2_x\,h(u)+h_1(u) u_x}=0,\label{cond2}
\end{eqnarray}
we get 
\begin{eqnarray}
h_1(u)=h_0
\end{eqnarray}
where $ h_0 $ is a constant.

We are able, now, to affirm:
\newtheorem {theorem}{Theorem}
\begin{theorem}
An equation belonging to the 
class (\ref{eq1}) can be transformed in a linear equation of the form
(\ref{linear}) by an ET generated by (\ref{gen}) if and only if the function $ f $  is given by
\begin{eqnarray}
f=u^2_x\,h(u)+h_0 u_x. \label{formaf}
\end{eqnarray}
\end{theorem}
{\it Proof.} From the eqns. (\ref{cond1}) and (\ref{cond2}) it follows that
the condition (\ref{formaf}) is necessary.

In order to demonstrate that it is sufficient, we must show that it exists at least a ET transforming the equations
\begin{eqnarray}
u_t-u_{xx}=u^2_x\,h(u)+h_0 u_x  \label{eqnonlinear}
\end{eqnarray}
in (\ref{linear}).
\par
The finite form of the  ETs generated   by (\ref{gen})
 is:
\begin{eqnarray}
t =\tau \,e^{-\varepsilon _1}-\varepsilon _0,\quad
 x =(\sigma -\tau \varepsilon _2-\varepsilon _3)e^{-\frac{1}{2}\varepsilon _1}, \quad
u =\psi (v) ,\label{trasformazione}
\end{eqnarray}
where $\psi $ is an arbitrary function, with $\psi  '(v)\not= 0$, and $\varepsilon_i$ are arbitrary parameters.

By applying the trasformation  (\ref{trasformazione}) 
to the equations (\ref{eqnonlinear}),
we get 
\begin{eqnarray}
v_\tau  -v_{\sigma \sigma }=
v^2_\sigma \,{\displaystyle\frac{{\psi  '}^2\,h(\psi  (v))+\psi  '' }{\psi  '}}
+\left(h_0e^{-\frac{1}{2}\varepsilon _1}-\varepsilon _2\right) v_\sigma .  \label{eqtrasf}
\end{eqnarray}
By choosing as $ \psi (v) $ a solution of ODE
\begin{eqnarray}
{\displaystyle\frac{{\psi  '}^2\,h(\psi  (v))+\psi  '' }{\psi  '}}=0, \label{eqphi}
\end{eqnarray}
the transformed equation (\ref {eqtrasf}) takes the linear form (\ref {linear})
where $k_0=h_0e^{-\frac{1}{2}\varepsilon _1}-\varepsilon _2.$  $\Box$ 

It is  a simple matter to show that it is possible to choose the arbitrary parameters $\varepsilon _i$
in order to make $k_0=0$.
So we can affirm, by assuming, for sake of simplicity, 
$\varepsilon _1=0$ and $\varepsilon _2=h_0$:
\newpage
\newtheorem {Corollary}{Corallary}
\begin{Corollary}
The group of ETs
\begin{eqnarray}
t =\tau-\varepsilon _0, \quad
 x =\sigma -\tau h_0-\varepsilon _3 ,\quad
u =H^{-1}(c_0 v+c_1), \label{trasformazionespec}
\end{eqnarray}
with $H^{-1}$ denoting the inverse function of
$H(\psi )=\int^{\psi }{e^{\int^{w}{h(z)dz}}dw}$ and  with $c_0\not= 0$, $c_1$ arbitrary constants,
 maps the equations of the form (\ref {eqnonlinear}) in the equation
\begin{eqnarray}
v_\tau  -v_{\sigma \sigma }=0. \nonumber
\end{eqnarray}
\end{Corollary}

\noindent
{\bf Example 1}
We consider the equation
\begin{eqnarray}
u_t -u_{xx}=-u^2_x\,tg\, u+u_x.  \label{eqesempio}
\end{eqnarray}
In this case is $h(u)=-tg\,u$ and $h_0=1$, so $H(\psi )=sin\,\psi $ and
the transformations (\ref {trasformazionespec}) become
\begin{eqnarray}
t =\tau-\varepsilon _0,\quad
 x =\sigma -\tau -\varepsilon _3 ,\quad
u =arcsin(c_0 v+c_1).\label{trasformazioneesempio}
\end{eqnarray}

\noindent  It is simple matter to verify that the transformation (\ref {trasformazioneesempio})
maps equation  (\ref {eqesempio}) in 
$$
v_\tau -v_{\sigma \sigma}=0. 
$$

\noindent
 \textbf{Remark}
The standard deterministic KPZ equation has the form
\begin{eqnarray}
h_t-Dh_{zz}=\lambda h_z^2
\end{eqnarray}
where $h(t,z)$ is the height of the surface at time $t$ above the point $z$ in the reference plane.
$Dh_{zz}$ describes diffusional relaxation within the surface. $D$ is the diffusion coefficient.
The strength of the nonlinearity $\lambda $ is proportional to the growth speed.

The above equation, after a trivial change of independent
variables $t=t$ and $ z=\sqrt{D}x$, reads
\begin{eqnarray}
h_t-h_{xx}=\frac{\lambda}{D} h_x^2. \label{KPZ}
\end{eqnarray}
A special case of this equation is the Burger's equation in potential form
\begin{eqnarray}
u_t-u_{xx}=u_x^2.\label{burger}
\end{eqnarray}
One can ascertain that the transformation 
(\ref{trasformazionespec}.III) for (\ref{KPZ}) and
 (\ref{burger}) becomes the well known  transformation
wich maps the considered equations in the well studied linear Fourier's equation
\begin{eqnarray}
w_t-w_{xx}=0.
\end{eqnarray}

\section{Conclusions}
\setcounter{equation}{0}
In this paper we considered a family of semilinear diffusion equations and
following \cite{NI}, \cite{NI1} we have obtained the differential invariants of second order
under the equivalence transformations for this family by the infinitesimal method.

As an application, we have proved that a family of generalized diffusion equations can be
reduced to the heat equation
\begin{eqnarray}
v_\tau -v_{\sigma \sigma }=0
\end{eqnarray}
via appropriate  equivalence transformations.

Finally, for special equations as standard deterministic  KPZ (Burger's equation in potential form),
from the transformation  (\ref {trasformazionespec})
we recovery the well known  transformation which brings them to the
heat equation.

\vskip 8mm
\noindent
 \textbf{Acknowledgements}

This work was supported by INdAM through G.N.F.M., by the M.I.U.R. project:
{\it Nonlinear mathematical problems of wave propagation and stability
in models of continuous media} and by University of Catania
(ex fondi 60\%).\\

\end{document}